\documentclass[a4paper,11pt]{article}
\pdfoutput=1 

\usepackage{jheppub} 
\usepackage[T1]{fontenc} 
\usepackage{esvect,bm,mathrsfs,makecell,graphicx,subcaption}
\usepackage{booktabs}
\usepackage[utf8]{inputenc}
\usepackage{slashed}
\usepackage{multirow}
\usepackage{color}
\newcommand{\wang}[1]{{\color{black} #1}}
\newcommand\ci{\perp\!\!\!\perp}

\title{The weak $B$, $B_s$ and $B_c$ decays to radially excited states}

\author{Tian Zhou$^a$\footnote{tianzhou@hit.edu.cn},}
\author{Tianhong Wang$^a$\footnote{thwang@hit.edu.cn (Corresponding author)},}
\author{Yue Jiang$^a$\footnote{jiangure@hit.edu.cn},}
\author{Lei Huo$^a$\footnote{lhuo@hit.edu.cn},}
\author{Guo-Li Wang$^{a,b,c}$\footnote{gl\_wang@hit.edu.cn},}
\affiliation{$^a$School of Physics, Harbin Institute of Technology, Harbin, 150001, China\\
$^b$Department of Physics, Hebei University, Baoding 071002, China\\
$^c$Hebei Key Laboratory of High-precision Computation and Application of Quantum Field Theory, Baoding
071002, China}

\abstract{Recently, many new excited states of heavy mesons have been discovered, especially radially excited states. It draws interest to study the production processes of these states from the ground $b$-flavored mesons. In this paper, we use the improved Bethe-Salpeter method to study the semi-leptonic and non-leptonic decays of $B$, $B_s$, and $B_c$ mesons. The calculations mostly focus on the decay channels with the radially excited $2S$ and $3S$ final states. We find that many channels have branching ratios up to $10^{-4}$, which are within the detection accuracy of current experiments.}

\begin{document}
\maketitle
\flushbottom


\section{Introduction}

There are plenty of experimental data of $b$-flavored heavy mesons, which have attracted a lot of attention. The $B$, $B_s$ and $B_c$ mesons, which are the ground states, have no electromagnetic and strong decay channels, so the two-body non-leptonic and three-body semi-leptonic decays play a central role in the research of these states. Due to the heavy masses, these mesons have very rich decay modes and comparatively long lifetimes \cite{Tanabashi:2018oca}. 

Nowadays, there are a lot of experiments studying the semi-leptonic decays of $B$, $B_s$ and $B_c$ mesons, for example, the results of BaBar \cite{Lees:2012xj}, Belle \cite{Sato:2016svk}, and LHCb \cite{Aaij:2017uff}. Similarly, there are also many experimental detections of the non-leptonic decays of these mesons, for example, the results of BaBar~\cite{581BARBAR:BDpi}, Belle~\cite{699Belle:B+DK-}, and LHCb~\cite{744LHCb:BsDsKpi}.

In addition to experiments, many theoretical and phenomenological methods have been used to study the decay caused by the $b \rightarrow c$ transition, for example, heavy quark sum rule (HQS) \cite{Neubert:1997uc}, QCD factorization \cite{Beneke:2000ry}, perturbative QCD (pQCD) \cite{Li:2008ts}, covariant light-front quark model (CLFQM) \cite{Li:2010bb}, light-cone QCD sum rules (LCSR) \cite{Li:2009wq}, Bethe-Salpeter (BS) equation method \cite{Chang:1992pt,Chang:2014jca}, Light-Front ISGW Model \cite{Anisimov:1998xv}, relativistic quark model \cite{Ebert:2003cn,Ivanov:2005fd,Ivanov:2000aj,Ivanov:2006ni}, and relativistic constituent quark model based on the Bethe-Salpeter formalism \cite{Liu:1997hr}.

Though there are a lot of experimental and theoretical studies on the $B$, $B_s$ and $B_c$ mesons decays, most of them only considered the case where the final meson is in the ground state, not the case where it is excited. Nowadays many excited states have been found. \wang{For example, except $\psi(2S)$ found years ago, $\eta_c(2S)$ has been discovered in Belle~\cite{Choi:2002na} decades ago.} A number of charmonium-like states have also been discovered in recent years. 
In the heavy-light sectors, many $2S$-state candidates have been found. The $D_{s1}^{*}(2710)^\pm$ discovered by Belle~\cite{Brodzicka:2007aa} has spin-parity $1^-$.  Theoretically, it was considered to be $D_s^*(2S)$~\cite{Matsuki:2006rz}, or $D_s^*(1D)$~\cite{Godfrey:2013aaa}, or mixture of them~\cite{Ebert:2009ua,Wang:2013mml}. In this paper, we assume it to be $D_s^*(2S)$. The $D(2550)^0$ discovered by LHCb~\cite{Aaij:2019sqk} has mass close to the theoretical prediction of $D(2S)^0$~\cite{Godfrey:1985xj}. Besides, LHCb has found another unnatural state $D_J(2580)^0$~\cite{Aaij:2013sza}, which may also be a candidate of $D(2S)^0$. The $D^*(2640)^\pm$ discovered by Delphi~\cite{Abreu:1998vk} have masses consistent with the predictions of $D^*(2S)^\pm$ in Ref.~\cite{Godfrey:1985xj}. \wang{The $D^*(2S)^0$ have many candidates, the $D_1^0(2680)^0$ observed by LHCb~\cite{Aaij:2016fma} in Dalitz plot analysis of $B^- \to D^+\pi\pi^-$ and the $D^*(2650)^0$ discovered by LHCb~\cite{Aaij:2013sza} may be the same states of $D^*(2S)^0$. Recently, LHCb also found the $D_1^*(2600)^0$ states~\cite{Aaij:2019sqk}, which might be the same object as the $D^*(2650)^0$ and $D_1^0(2680)^0$. There is still some tension in the measurements of the parameters, but the spin-parity of $D_1^0(2680)^0$ and $D_1^*(2600)^0$ is confirmed as $1^-$. In this paper, we assume the $D_1^*(2680)^0$ as the $D^*(2S)^0$. Sadly, there are still no candidates for the $D(2S)^+$ or $D_s(2S)^+$ states have yet been observed up to now.}

In a previous paper~\cite{Geng:2018qrl}, we find that the contribution of the relativistic correction cannot be ignored, especially when excited meson are included. Therefore, in the research involving excited states, relativistic methods are needed. In this paper, we use the improved BS method which has been applied in our previous paper~\cite{Zhou:2019stx}.

The BS equation relativistically describes two-body bound states. By making an instantaneous approximation for the interaction kernel, the BS equation is reduced to the full Salpeter equation, which can be solved numerically to obtain the wave functions of different states. These wave functions imply relativistic corrections. The next step is to calculate the transition matrix elements which can be written as the overlap integrals of the wave functions of the initial and final mesons. Part of the relativistic corrections are included by boosting the wave functions of the final meson from its rest frame to the moving one~\cite{Fu:2011tn, Zhou:2019stx}. 

The paper is organized as follows. In Section 2, we present the formula of the transition matrix element by the improved BS method and show the definitions of form factors for different decay channels. In Section 3, we use these form factors to calculate the non-leptonic decay processes of ground $b$-flavored mesons. In Section 4, the numerical results and discussions are presented.


\section{Formalism of semi-leptonic decays}

\begin{figure}
  \centering
  \includegraphics[scale=0.40]{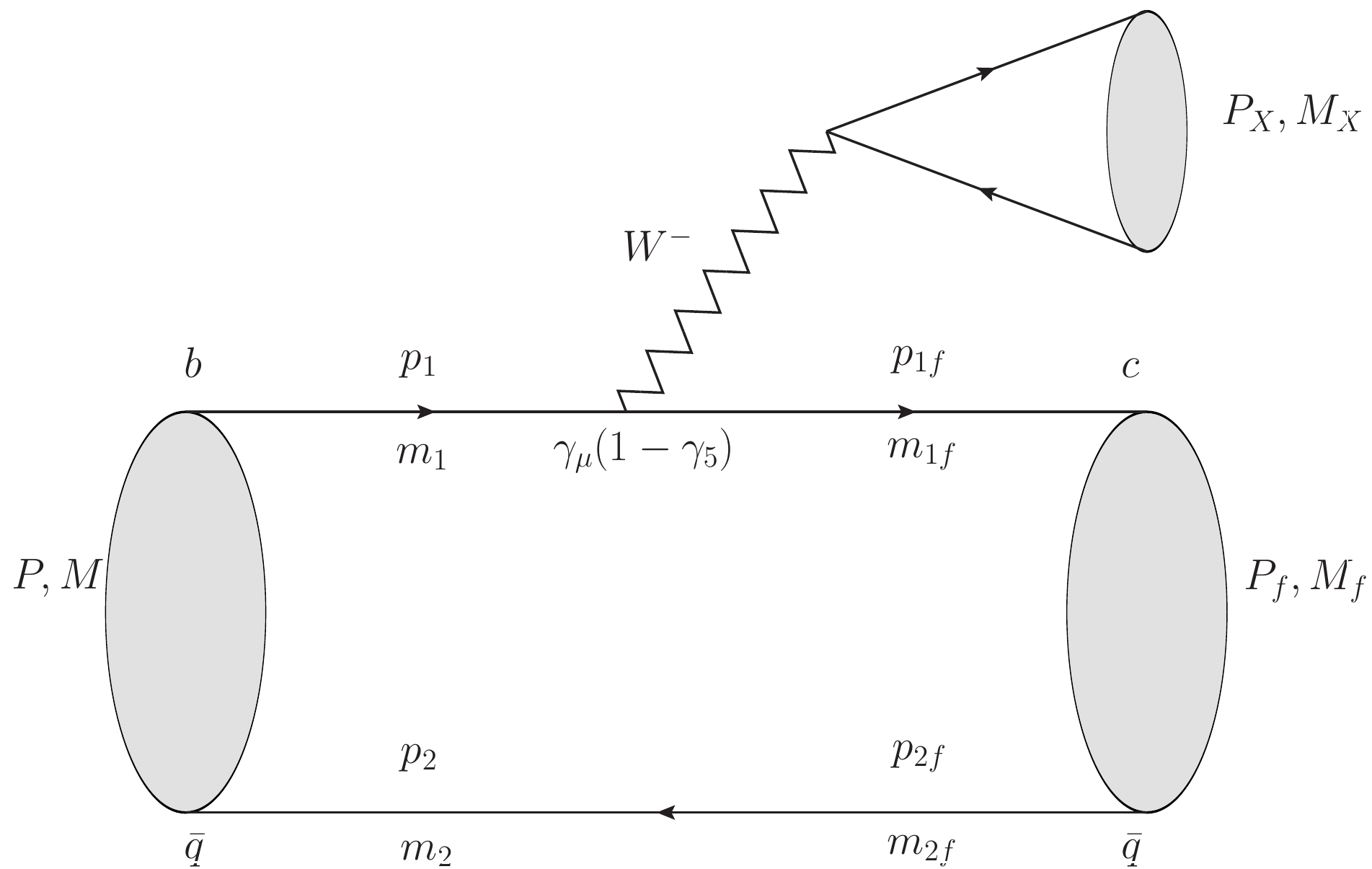}
  \caption{\label{fig:b-semi} Feynman diagram of the semi-leptonic decay of $B_q$ to $D_q$ ($q=u,d,s,c$), where $P(M)$ and $P_f(M_f)$ are the momenta (masses) of $B_q$ and $D_q$, respectively; $m_i(m_{if})$ and $p_i(p_{if})$ are respectively the masses and momenta of the quark or antiquark in the initial (final) state.}
\end{figure}

In this section, we present the formula of semi-leptonic transitions of $B_q$ ($q=u,d,s,c$) to $D_q$ with the improved BS method. Fig.\ref{fig:b-semi} is the Feynman diagram responsible for such decay processes whose amplitudes have the form
\begin{equation}
\label{e1}
\begin{split}
T=\frac{G_F}{\sqrt{2}} V_{bc}\bar{u}_{\ell}\gamma^\mu(1-\gamma_5)v_{\bar{\nu}_\ell} \langle D_q|J_\mu|B_q\rangle.
\end{split}
\end{equation}
According to the Mandelstam formalism, the transition matrix element $\langle D_q|J_\mu|B_q\rangle$ can be written as the overlap integral over initial and final states wave functions:
\begin{equation}\label{eq:amporigin}
\left\langle D_q \left(P_{f}\right)\right|(\overline{c}\Gamma^{\mu} b )|B_q (P)\rangle=\int \frac{\mathrm{d}^{4} q}{(2 \pi)^{4}} \frac{\mathrm{d}^{4} q_{f}}{(2 \pi)^{4}} \operatorname{Tr}\left[\overline{\chi}_{P_{f}}\left(q_{f}\right) \Gamma^{\mu} \chi_{P}(q) \mathrm{i} S_{2}^{-1}\left(p_{2}\right)\right](2 \pi)^{4} \delta^{4}\left(p_{2}-p_{2 f}\right),
\end{equation}
where we have defined $\Gamma^\mu=\gamma^\mu(1-\gamma^5)$. 

As it is hard to solve the four-dimensional BS equation, we will reduce the equation to its instantaneous version, namely the three dimensional Salpeter equation. Then, with some approximations to the numerator of the integrand~\cite{Zhou:2019stx}, the transition matrix can be written finally as,
\begin{equation}
\begin{aligned}
\left\langle D_q \left(P_{f}\right)\right|(\overline{c}\Gamma^{\mu} b )|B_q (P)\rangle=-i\int\frac{d\vec q}{(2\pi)^3}{\rm Tr}\Big[\frac{\slashed P_f}{M_f}\overline\varphi^{++}(P_f, q_{f\ci})\frac{\slashed P_f}{M_f}L_{r}\Gamma^\mu \varphi^{++}(P, q_\perp)\Big],
\end{aligned}
\end{equation}
where we have defined
\begin{equation}
L_{r}=\frac{\left(M_{f}-\widetilde\omega_{1f}-\widetilde\omega_{2f}\right)}{\left(P_{f_{P}}-\omega_{1 f}-\omega_2\right)} \Lambda_{1}^{+}\left(p_{1f\perp}\right),
\end{equation}
and the positive energy part of the wave functions 
\begin{equation}
  \begin{aligned}
  &\varphi^{++}(P,q_\perp) = \Lambda_1^{+}(p_{1\perp})\frac{\slashed P}{M}\varphi(P, q_\perp)\frac{\slashed P}{M}\Lambda_2^{+}(p_{2\perp}),\\
  &\varphi^{++}(P_f,q_{f\ci}) = \widetilde\Lambda_1^{+}(p_{1f\ci})\frac{\slashed P_f}{M_f}\varphi(P_f, q_{f\ci})\frac{\slashed P_f}{M_f}\widetilde\Lambda_2^{+}(p_{2f\ci}),
  \end{aligned}
  \end{equation} 
whose explicit forms can be found in our previous paper\cite{Zhou:2019stx}.  

The symbols used in the above equations are illustrated as follows. $\widetilde\omega_{j f}$ and $\omega_{j}$ are defined as
\begin{equation}
  \begin{aligned}
  \widetilde\omega_{if}\equiv\sqrt{m_{if}^2-p_{if\ci}^2}=\sqrt{m_{if}^2-q_{f\ci}^2},\\
  \omega_{if}\equiv\sqrt{m_{if}^2-p_{if\perp}^2}=\sqrt{m_{if}^2-q_{f\perp}^2},
  \end{aligned}
\end{equation}
where we have used $\ci$ and $\perp$ to represent projecting to the momenta of the final and initial mesons, respectively. $q_{f\ci}$ and $q_{f\perp}$ are related by 
\begin{equation}
q_{f\ci}=q_{f\perp}-\frac{q_{f\perp} \cdot P_{f\perp}}{M_{f}^{2}} P_{f}+s_{r}\left(\frac{1}{M} P-\frac{P_{f_{_{P}}}}{M_{f}^{2}} P_{f}\right),
\end{equation}
where $s_{r}=\alpha_{2} P_{f_{p}}-\omega_{2 f_{P}}$. The projection operators $\widetilde\Lambda^\pm_i$ are defined as
\begin{equation}
  \widetilde\Lambda^\pm_i(p_{if\ci})=\frac{1}{2\widetilde\omega_{if}}\Big[\frac{\slashed P_f}{M_f}\widetilde\omega_{if} \pm(Jm_{if}+\slashed p_{if\ci})\Big],
\end{equation}
with $J=1$ and $-1$ for the cases quark and anti-quark, respectively.

In this paper, we focus on the cases where the final excited state is a pseudoscalar or a vector, then the corresponding form factors can be defined as functions of momentum transfer, $Q^\mu=P^\mu-P_f^\mu$. For the $B_q \rightarrow D_q^{(*)}$ process, the form factors corresponding to the vector and axial currents are defined as 
\begin{equation}\label{eq:formfactor1}
\left\langle D\left|\bar{c} \gamma^{\mu} b\right| \bar{B}\right\rangle=F_{+}\left(Q^{2}\right)\left[\left(P+P_f\right)^{\mu}-\frac{m_{B}^{2}-m_{D}^{2}}{Q^{2}} Q^{\mu}\right]+F_{0}\left(Q^{2}\right) \frac{m_{B}^{2}-m_{D}^{2}}{Q^{2}} Q^{\mu},
\end{equation}
\begin{equation}\label{eq:formfactor2}
\left\langle D^{*}\left|\bar{c} \gamma^{\mu} b\right| \bar{B}\right\rangle=\frac{2 i V\left(Q^{2}\right)}{m_{B}+m_{D^{*}}} \epsilon^{\mu \nu \rho \sigma} P_{f\nu} P_{\rho} \varepsilon_{\sigma},
\end{equation}
\begin{equation}\label{eq:formfactor3}
  \begin{aligned}
  \left\langle D^{*}\left|\bar{c} \gamma^{\mu} \gamma_{5} b\right| \bar{B}\right\rangle=\left(m_{B}+m_{D^{*}}\right) A_{1}\left(Q^{2}\right) \varepsilon^{ \mu}-A_{2}\left(Q^{2}\right) \frac{\varepsilon \cdot Q}{m_{B}+m_{D^{*}}}\left(P+P_f\right)^{\mu} \\
  \quad -\frac{\varepsilon \cdot Q}{Q^{2}} Q^{\mu}\left[\left(m_{B}+m_{D^{*}}\right) A_{1}\left(Q^{2}\right)-\left(m_{B}-m_{D^{*}}\right) A_{2}\left(Q^{2}\right)-2 m_{D^{*}} A_{0}\left(Q^{2}\right)\right],
  \end{aligned}
\end{equation}
where $\varepsilon$ is the polarization vector of the $D^\ast$ meson.


\section{Formalism of non-leptonic decays}

In this section, we will present the formula of a $B_q$ $(q=u,d,s,c)$ meson non-leptonic decays to a heavy charmed meson and a light meson $X$ $(X=\rho, \pi, K, K^*)$ with the improved BS method. The effective Lagrangian which describes such processes can be written as \cite{Neubert:1997uc}
\begin{equation}
\begin{aligned} 
H_{\mathrm{eff}}= \frac{G_{F}}{\sqrt{2}}\left\{\sum_{i=1}^{2} c_{i}(\mu)\left[V_{u b} V_{u s}^{*} O_{i}^{u}(\mu)+V_{c b} V_{c s}^{*} O_{i}^{c}(\mu)\right]+\mathrm{h.c.}\right\}.
\end{aligned}
\end{equation}
Here we have used the definitions of the effective operators
\begin{equation}\begin{array}{ll}
  O_{1}^{u}=\left(\bar{s}_{\alpha} u_{\beta}\right)_{V-A}\left(\bar{u}_{\beta} b_{\alpha}\right)_{V-A}, & O_{2}^{u}=(\bar{s} u)_{V-A}(\bar{u} b)_{V-A},\\
  O_{1}^{c}=\left(\bar{s}_{\alpha} c_{\beta}\right)_{V-A}\left(\bar{c}_{\beta} b_{\alpha}\right)_{V-A}, & O_{2}^{c}=(\bar{s} c)_{V-A}(\bar{c} b)_{V-A},
  \end{array}
\end{equation}
where $(\bar{q_1} q_2)_{V-A}=\bar{q_1}\gamma_{\mu}\left(1-\gamma_{5}\right) q_2$; $\alpha$, $\beta$ are the  color indices. 

In the calculation, we adopt the combinations $c_{\pm}(\mu)=c_{1}(\mu) \pm c_{2}(\mu)$ of the Wilson coefficients. The evolution of $c_\pm(\mu)$ can be obtained by the QCD Renormalization Group Equation (RGE), which reads \cite{Buchalla:1995vs}
\begin{equation}\label{eq:RGE}
\left(\mu \frac{\mathrm{d}}{\mathrm{d} \mu}-\Gamma_{\pm}\right) c_{\pm}(\mu)=0,
\end{equation}
where at one-loop order
\begin{equation}
\Gamma_{\pm}=\gamma_{\pm} \frac{\alpha_{s}}{4 \pi} ; \quad \gamma_{\pm}=6\left(\pm 1-\frac{1}{N_{c}}\right).
\end{equation}
The running coupling constant $\alpha_s(\mu)$ can be written as \cite{Buchalla:1995vs}
\begin{equation}
\alpha_{s}(\mu)=\frac{4 \pi}{\beta_{0} \ln \left(\mu^{2} / \Lambda_{\mathrm{QCD}}^{2}\right)}\left[1-\frac{\beta_{1}}{\beta_{0}^{2}} \frac{\ln \ln \left(\mu^{2} / \Lambda_{\mathrm{QCD}}^{2}\right)}{\ln \left(\mu^{2} / \Lambda_{\mathrm{QCD}}^{2}\right)}\right],
\end{equation}
where
\begin{equation}
\beta_{0}=\frac{11 N_{c}-2 f}{3}, \quad \beta_{1}=\frac{34}{3} N_{c}^{2}-\frac{10}{3} N_{c} f-2 C_{F} f, \quad C_{F}=\frac{N_{c}^{2}-1}{2 N_{c}}.
\end{equation}
The parameters related above are:  the number of colors $N_{c} = 3$,  the number of light flavors $f=5$, $\Lambda_{\mathrm{QCD}}=0.27$ GeV. Then the solution of Eq.\eqref{eq:RGE} at leading order can be written as \cite{Buchalla:1995vs}
\begin{equation}\label{eq:wilsion}
c_{\pm}(\mu)=\left[\frac{\alpha_{s}\left(M_{W}\right)}{\alpha_{s}(\mu)}\right]^{\gamma_{\pm}/2\beta_0},
\end{equation}
where $M_W=80.379$ GeV is used.

To calculate the decay widths, we will change the forms of the form factors we presented in the Section 2. Their expressions can be found in our previous paper \cite{Zhou:2019stx}. Using the definitions of form factors in Eq.\eqref{eq:formfactor1}, Eq.\eqref{eq:formfactor2} and Eq.\eqref{eq:formfactor3}, we can calculate the decay amplitudes within the naive factorization scheme, which can be written as \cite{Beneke:2000ry}
\begin{equation}
\begin{aligned} \mathcal{A}\left(\bar B^0 \rightarrow D^{+} \pi^{-}\right) &=i \frac{G_{F}}{\sqrt{2}} V_{u d}^{*} V_{c b} a_{1} f_{\pi} F_{0}\left(m_{\pi}^{2}\right)\left(m_{B}^{2}-m_{D}^{2}\right), \\ \mathcal{A}\left(\bar B^0 \rightarrow D^{*+} \pi^{-}\right) &=-i \frac{G_{F}}{\sqrt{2}} V_{u d}^{*} V_{c b} a_{1} f_{\pi} A_{0}\left(m_{\pi}^{2}\right) 2 m_{D} \varepsilon^{*} \cdot p, \\ \mathcal{A}\left(\bar B^0 \rightarrow D^{+} \rho^{-}\right) &=-i \frac{G_{F}}{\sqrt{2}} V_{u d}^{*} V_{c b} a_{1} f_{\rho} F_{+}\left(m_{\rho}^{2}\right) 2 m_{\rho} \eta^{*} \cdot p, \\ \mathcal{A}\left(\bar B^0 \rightarrow D^{*+} \rho^{-}\right)=& i \frac{G_{F}}{\sqrt{2}} V_{u d}^{*} V_{c \delta} \varepsilon^{* \mu} \eta^{* \nu}\left(S_{1} g_{\mu \nu}-S_{2} q_{\mu} p_{\nu}^{\prime}+i S_{3} \epsilon_{\mu \nu \alpha \beta} p^{\prime \alpha} q^{\beta}\right), \end{aligned}
\end{equation} 
where $f_{\pi,\rho}$ are the decay constants and $\eta^\ast$ is the polarization vector of $\rho$ meson. We also use the following definitions
\begin{equation}
\begin{aligned}
|\vec{q}|&=\frac{1}{2 m_{B}} \sqrt{\left(m_{B}^{2}-m_{1}^{2}-m_{2}^{2}\right)^{2}-4 m_{1}^{2} m_{2}^{2}}, \\
 S_{1}&=a_{1}\left(D^{*} \rho\right) m_{\rho} f_{\rho}\left(m_{B}+m_{D^{*}}\right) A_{1}\left(m_{\rho}^{2}\right), \\
  S_{2}&=a_{1}\left(D^{*} \rho\right) m_{\rho} f_{\rho} \frac{2 A_{2}\left(m_{\rho}^{2}\right)}{m_{B}+m_{D}}.
\end{aligned}
\end{equation}

We definine the following functions
\begin{equation}
\begin{aligned} H_{0} &=\frac{1}{2 m_{D^{*}} m_{\rho}}\left[\left(m_{B}^{2}-m_{D^{*}}^{2}-m_{\rho}^{2}\right) S_{1}-2 m_{B}^{2}|\vec{q}|^{2} S_{2}\right], \\ H_{ \pm} &=S_{1} \mp m_{B}|\vec{q}| S_{3},
 \end{aligned}
\end{equation}
where, at the leading order, $a_1$ can be written as \cite{Beneke:2000ry}
\begin{equation}
a_{1}=c_{1}(\mu)+\xi c_{2}(\mu), \quad \xi=\frac{1}{N_{c}}.
\end{equation}
Then we get the partial widths of the non-leptonic decays, which can be written as
\begin{equation}\label{eq:final}
\begin{aligned} \Gamma\left(\bar B^0 \rightarrow D^{+} \pi^{-}\right) &=\frac{G_{F}^{2}\left(m_{B}^{2}-m_{D}^{2}\right)^{2}|\vec{q}|}{16 \pi m_{B}^{2}}\left|V_{u d}^{*} V_{c b}\right|^{2}\left|a_{1}\right|^{2} f_{\pi}^{2} F_{0}^{2}\left(m_{\pi}^{2}\right), \\ \Gamma\left(\bar B^0 \rightarrow D^{*+} \pi^{-}\right) &=\frac{G_{F}^{2}|\vec{q}|^{3}}{4 \pi}\left|V_{u d}^{*} V_{c b}\right|^{2}\left|a_{1}\right|^{2} f_{\pi}^{2} A_{0}^{2}\left(m_{\pi}^{2}\right), \\ \Gamma\left(\bar B^0 \rightarrow D^{+} \rho^{-}\right) &=\frac{G_{F}^{2}|\vec{q}|^{3}}{4 \pi}\left|V_{u d}^{*} V_{c b}\right|^{2}\left|a_{1}\right|^{2} f_{\rho}^{2} F_{+}^{2}\left(m_{\rho}^{2}\right), \\ \Gamma\left(\bar B^0 \rightarrow D^{*+} \rho^{-}\right) &=\frac{G_{F}^{2}|\vec{q}|}{16 \pi m_{B}^{2}}\left|V_{u d}^{*} V_{c b}\right|^{2}\left(\left|H_{0}\right|^{2}+\left|H_{+}\right|^{2}+\left|H_{-}\right|^{2}\right).
\end{aligned}
\end{equation}
When the final light meson is $K$ or $K^*$, the calculation is similar to that of the $\pi$ or $\rho$ case, respectively.


\section{Numerical Results and Discussions}

In this work, we choose the Cornell potential as the interaction kernel \cite{Kim:2003ny}, which is the addition of a linear and a vector potential. In the momentum space, it has the form
\begin{equation}
\label{ea12}
\begin{split}
V(\vec{q}) & =V_s(\vec{q})+V_v(\vec{q})\gamma_0\otimes\gamma^0,
\\
V_s(\vec{q}) & =-\left( \frac{\lambda}{\alpha} +V_0 \right) \delta^3(\vec{q}) + \frac{\lambda}{\pi^2}\frac{1}{(\vec{q}^2+\alpha^2)^2},
\\
V_v(\vec{q}) & =-\frac{2}{3\pi^2}\frac{\alpha_s(\vec{q})}{(\vec{q}^2+\alpha^2)},
\end{split}
\end{equation}
where the symbol $\bigotimes$ denotes that the Salpeter wave function is sandwiched between the two $\gamma^0$ matrices; the QCD running coupling constant $\alpha_s(\vec{q})$ has the expression $\alpha_s(\vec{q})=\frac{12\pi}{33-2N_f}\frac{1}{\mathtt{ln}\left( a+\vec{q}^2/\Lambda^2_{QCD} \right)}$. To solve the Salpeter equations, we have to fix the relevant parameters by fitting the mass spectra of mesons. Here we choose the same values as in those in Ref.~\cite{Wang:2012cp}
\begin{equation}\label{parameter}
\begin{array}{llr}{a=e=2.7183,} & {\alpha=0.060 ~\rm{GeV},} & {\lambda=0.210~ \rm{GeV}^{2}}, \\ {m_{u}=0.305 ~\rm{GeV},} & {m_{d}=0.311 ~\rm{GeV},} & {m_{s}=0.500~ \rm{GeV}}, \\ {m_{c}=1.62 ~\rm{GeV},} & {m_{b}=4.96~ \rm{GeV},} & {\Lambda_{\rm{QCD}}=0.270 ~ \rm{GeV}}.\end{array}
\end{equation}
The CKM matrix elements and the decay constants, which are used to calculate the transition matrix element, are taken from PDG~\cite{Tanabashi:2018oca}:
 $|V_{c b}|=0.0411$, $|V_{u s}|=0.225$, $|V_{c s}|=0.9735$, $f_{\pi} =0.130$~GeV, $f_{\rho} =0.205 \pm 0.009$~GeV, $f_{K} =0.156$~GeV, $f_{K^{*}} =0.217 \pm 0.005$~GeV.

The full Salpeter equations fulfilled by the wave functions of the $0^-$ and $1^-$ mesons have been solved in Ref.~\cite{Kim:2003ny} and Ref.~\cite{Wang:2005qx}, respectively. Here, we will not show the details, but present the masses of the excited states in table \ref{tab:mass and decay constant in excited}. The numerical values of the wave functions are used to calculate the partial widths of the semi-leptonic and non-leptonic decay channels.

\begin{table}[htb]
  \caption{ The masses (MeV) of mesons in the excited states.}
   \label{tab:mass and decay constant in excited}
   \centering
   \begin{tabular}{cccccccccc}
    \Xhline{1.5pt}
        particle       & Mine       & Exp. & Godfrey~\cite{Godfrey:2015dva,Barnes:2005pb} & Ebert~\cite{Ebert:2003cn} & Wang~\cite{Wang:2014lml} \\
    \hline
        $B^0(2S)$      & 5840       &                                        &                                      & 5890                &                    \\
        $B_s^0(2S)$    & 5930       &                                        &                                      & 5976                &                    \\
        $D^0(2S)$      & 2550       & \wang{$2518 \pm 2 \pm 7$~\cite{Aaij:2019sqk}}              & 2581               & 2581               &                    \\
        $D^+(2S)$      & 2555       &                & 2581                                 & 2581               &                    \\
        $D_s^+(2S)$    & 2670       &                                        & 2673                                 & 2688                &                    \\
        $\eta_c(2S)$   & 3638       & $3637.5 \pm 1.1$~\cite{Tanabashi:2018oca}                       & 3630                                 &                     & 3622               \\
    \hline
        $B^{*0}(2S)$   & 5880       &                                        &                                      & 5906                &                    \\
        $B_s^{*0}(2S)$ & 5955       &                                        &                                      & 5992               &                    \\
        $D^{*0}(2S)$   & 2660       &    \wang{$2681.1\pm5.6\pm4.9\pm13.1$~\cite{Aaij:2016fma}}           & 2643                                & 2632                &                    \\
        $D^{*+}(2S)$   & 2605       &    $2637 \pm 2 \pm 6$~\cite{Abreu:1998vk}                  & 2643                                 & 2632                &                    \\
        $D_s^{*+}(2S)$ & 2710       & $2732 \pm 4.3 \pm 5.8$~\cite{Aaij:2016utb}     & 2732                                 & 2731               &                    \\
        $\psi(2S)$     & 3686       & $3686.10 \pm 0.06$~\cite{Tanabashi:2018oca}                     & 3672                                 &                     & 3684               \\
    \hline
        particle       & Mine       & Exp. & Godfrey~\cite{Godfrey:2015dva,Barnes:2005pb} & Ebert~\cite{Ebert:2003cn} & Wang~\cite{Wang:2014lml} \\
    \hline
        $D^0(3S)$      & 3030       &                                        & 3068                                 & 3062 \\
        $D^+(3S)$      & 3035       &                                        & 3068                                 & 3062 \\
        $D_s^+(3S)$    & 3150       &                                        & 3154                                 & 3219 \\
        $\eta_c(3S)$   & 3990       &                                        & 4043                                 &                     & 4007 \\
    \hline
        $D^{*0}(3S)$   & 3080       &                                        & 3110                                 & 3096 \\
        $D^{*+}(3S)$   & 3085       &                                        & 3110                                & 3096 \\
        $D_s^{*+}(3S)$ & 3190       &                                        & 3193                                 & 3242 \\
        $\psi(3S)$     & 4040       & $4039 \pm 1$~\cite{Tanabashi:2018oca}                           &  \\
    \Xhline{1.5pt}
    \end{tabular}%
\end{table}%

In our previous work~\cite{Zhou:2019stx}, this improved BS method has been used to study the semi-leptonic decays of  $B$, $B_s$ and $B_c$ mesons, where the final states are ground charmed mesons. There, the results we got are consistent with the experimental data, which indicates that this method is validity to investigate the decay properties of heavy mesons.

In this work, we first calculate the non-leptonic decays of heavy mesons to $1S$ states and one light meson. By using Eq.~\eqref{eq:wilsion}, we calculate the Wilson coefficients at the leading order, and the results are listed in table \ref{tab:wilson}. The partial decay widths and branching fractions are presented in tables~\ref{tab:BD1sX} and \ref{tab:BDstar1sX}, which corresponding to the channels with final heavy meson being $0^-$ and $1^-$, respectively. The experimental data from PDG are also presented as a comparison. Besides, the results of the $B_c$ meson decaying to $1S$ states have been calculated in our previous work~\cite{Chang:2014jca} with the same method. So we will not present the results here. One can see that most theoretical predictions are consistent with data.

Besides, there are many experimental results for the ratios of branching fractions of the non-leptonic decays. As a check, we also calculate such ratios, which are presented in table \ref{tab:compare}. As the CKM matrix elements and the decays constants give no contribution, the theoretical values of such ratios can be used to compare with the experimental data, which may provide useful information of the form factors. We can see that within the experimental errors, our results agree well with experimental data, which confirms that the improved BS method is a good and suitable way to study such decays of heavy mesons. Next, we apply this method to study the weak decays of $B$, $B_s$, and $B_c$ mesons, where the final heavy meson are radially excited states.

\begin{table}[htbp]
\centering
 \caption{\label{tab:wilson}Values of the Wilson coefficients at the scale $m_b=4.96$~GeV and $m_c=1.62$~GeV.}
 \begin{tabular}{ccccc}
   \Xhline{1.5pt}
Mass   & $c_1^{LO}(m)$ & $c_2^{LO}(m)$ & $a_1=c1+c2/N_c$  \\
  \midrule
$m_b$ & 1.111  & -0.256  & 1.026   \\
$m_c$ & 1.241  & -0.479  & 1.082  \\
\bottomrule
 \end{tabular}
\end{table}

\begin{table}[htbp]
\centering
 \caption{\label{tab:BD1sX} Branching ratios ($10^{-3}$) and partial widths ($10^{-15}$ GeV) of $0^- \rightarrow 0^-(1S)X$.}
 \begin{tabular}{cccccccccccccc}
   \Xhline{1.5pt}
  $\bar{B}^0\rightarrow D^+(1S)X^-$        & width   &   Br      &   Br(Exp.) \cite{Tanabashi:2018oca}    \\
  \midrule
$\pi^-$                          & $1.40^{+0.22}_{-0.18}$       &  $3.24^{+0.50}_{-0.42}$              &  $2.52 \pm 0.13$ \\
$K^-$                             &$0.105^{+0.016}_{-0.013}$       &  $0.245^{+0.038}_{-0.031}$                &  $0.186\pm 0.020$  \\
$\rho^-$                           &$3.37^{+0.52}_{-0.43}$      &  $7.91^{+1.20}_{-1.01}$                   &$7.6 \pm 1.2$ \\
$K^{*-}$                         & $0.187^{+0.029}_{-0.024}$      &  $0.431^{+0.067}_{-0.056}$              & $0.45\pm 0.07$   \\
  \midrule
$B^-\rightarrow D^0(1S)X^-$           & width    &   Br      &   Br(Exp.) \cite{Tanabashi:2018oca}     \\
  \hline
$\pi^-$                            &$1.40^{+0.22}_{-0.18}$       &  $3.49^{+0.54}_{-0.45}$              &  $4.68\pm 0.13$  \\
$K^-$                               & $0.105^{+0.016}_{0.013}$       &  $0.264^{+0.040}_{-0.034}$                & $0.363\pm 0.012$  \\
$\rho^-$                            &$3.37^{+0.52}_{-0.43}$      &  $8.40^{+1.29}_{-1.09}$                 & $13.4 \pm 1.8$ \\
$K^{*-}$                           & $0.187^{+0.029}_{-0.024}$      &  $0.466^{+0.072}_{-0.060}$              & $0.53\pm 0.04$ \\
  \midrule
$\bar{B}^0_s\rightarrow D^+_s(1S)X^-$      & width   &   Br      &   Br(Exp.) \cite{Tanabashi:2018oca}     \\
  \hline
$\pi^-$                            &$1.27^{+0.22}_{-0.18}$       &  $2.92^{+0.50}_{-0.42}$              &  $3.00 \pm 0.23$  \\
$K^-$                              & $0.0961^{+0.0164}_{-0.0138}$       &  $0.221^{+0.038}_{-0.032}$                &  $0.227\pm0.019$   \\
$\rho^-$                           &$3.06^{+0.53}_{-0.44}$      &  $7.04^{+1.21}_{-1.01}$              &  $6.9 \pm 1.4$  \\
$K^{*-}$                         &$0.169^{+0.029}_{-0.024}$      &  $0.392^{+0.068}_{-0.056}$            \\
  \bottomrule
 \end{tabular}
\end{table}

\begin{table}[htbp]
\centering
 \caption{\label{tab:BDstar1sX} Branching ratios($10^{-3}$) and partial widths ($10^{-15}$ GeV) of  $0^- \rightarrow 1^-(1S)X$.} 
 \begin{tabular}{ccccccccccccc}
   \Xhline{1.5pt}
  $\bar{B}^0\rightarrow D^{*+}(1S)X^-$        & width  &   Br      &   Br(Exp.)  \cite{Tanabashi:2018oca} \\
  \midrule
$\pi^-$                           &$1.64^{+0.25}_{-0.21}$       &  $3.80^{+0.58}_{-0.49}$              &  $2.74\pm 0.13$  \\
$K^-$                              &$0.122^{+0.019}_{-0.016}$       &  $0.281^{+0.043}_{-0.036}$                &  $0.212 \pm 0.015$  \\
$\rho^-$                          & $3.45^{+0.31}_{-0.29}$      &  $8.73^{+0.78}_{-0.73}$                  & $6.8 \pm 0.9$\\
$K^{*-}$                          & $0.328^{+0.028}_{-0.026}$      &  $0.758^{+0.064}_{-0.059}$              & $0.33 \pm 0.06$  \\
  \midrule
$B^-\rightarrow D^{*0}(1S)X^-$           & width    &   Br      &   Br(Exp.) \cite{Tanabashi:2018oca}    \\
  \hline
$\pi^-$                            &$1.64^{+0.25}_{-0.21}$       & $4.11^{+0.62}_{-0.53}$              &  $4.90 \pm 0.17$  \\
$K^-$                              &$0.121^{+0.019}_{-0.016}$       &  $0.304^{+0.047}_{-0.039}$                &  $0.397^{+0.31}_{-0.28}$   \\
$\rho^-$                         & $3.49^{+0.32}_{-0.29}$      &  $8.73^{+0.79}_{-0.73}$                  & $9.8 \pm 1.7$\\
$K^{*-}$                          & $0.339^{+0.028}_{-0.026}$      &  $0.846^{+0.070}_{-0.064}$              & $0.81 \pm 0.14$   \\
  \midrule
$\bar{B}^0_s\rightarrow D^{*+}_s(1S)X^-$       & width   &   Br      &   Br(Exp.) \cite{Tanabashi:2018oca}   \\
  \hline
$\pi^-$                            &$1.46^{+0.25}_{-0.21}$       &  $3.37^{+0.57}_{-0.47}$              &  $2.0 \pm 0.5$   \\
$K^-$                             &$0.108^{+0.018}_{-0.015}$       & $0.249^{+0.042}_{-0.035}$                &  $0.133 \pm 0.35$  \\
$\rho^-$                           &$3.18^{+0.33}_{-0.31}$      &  $7.26^{+0.76}_{-0.71}$              & $9.6 \pm 2.1$  \\
$K^{*-}$                          &$0.299^{+0.029}_{-0.028}$      & $0.688^{+0.067}_{-0.064}$                \\
  \bottomrule
 \end{tabular}
\end{table}

\begin{table}[htbp]
\centering
\caption{\label{tab:compare} The ratios of branching fractions for the non-leptonic decays. The results indicated by `PDG' is achieved by taking the values of branching ratios from PDG.}
\begin{tabular}{ccccc}
  \Xhline{1.5pt}
  Parameter   &   Measurements   &   Improved BS
\\
\midrule
\multirow{2}{*}{$\frac{Br\left(B^{-} \rightarrow D^{*0} \pi^{-}\right) }{Br\left(B^{-} \rightarrow D^{0} \pi^{-}\right)}$ } &  BaBar~\cite{Aubert:2006jc}:$1.14 \pm 0.07 \pm 0.04$ & \multirow{2}{*}{1.17}
\\
& \wang{PDG~\cite{Tanabashi:2018oca}:$ 1.126^{+0.007}_{-0.008} $} &
\\[0.5cm]
\multirow{2}{*}{ $\frac{Br\left(B^{-} \rightarrow D^{*0} K^{-}\right) }{Br\left(B^{-} \rightarrow D^{0} K^{-}\right)}$ } & LHCb \cite{Aaij:2017ryw}:$0.992 \pm 0.077$  & \multirow{2}{*}{0.837}
\\
& \wang{PDG \cite{Tanabashi:2018oca}:$ 1.093^{+0.048}_{-0.042} $} &
\\[0.5cm]
\multirow{2}{*}{ $\frac{Br\left(B_s^{0} \rightarrow D_s^{-} \pi^{+}\right) }{Br\left(B^{0} \rightarrow D^{-} \pi^{+}\right)}$ } &  CDF \cite{Abulencia:2006aa}:$1.13\pm0.08\pm 0.23$ & \multirow{2}{*}{0.901}
\\
& \wang{PDG \cite{Tanabashi:2018oca}:$ 1.190^{+0.028}_{-0.031} $} &
\\[0.5cm]
$\frac{Br\left(B_{c}^{+} \rightarrow \psi(2S) \pi^{+}\right) }{Br\left(B_{c}^{+} \rightarrow J / \psi \pi^+\right)}$ & LHCb \cite{Aaij:2015xga}:$0.268 \pm 0.032 \pm 0.007\pm0.006$ & 0.208
\\
  \bottomrule
\end{tabular}
\end{table}

In tables \ref{tab:002s} and \ref{tab:003s}, we give the results of heavy bottom mesons semi-leptonic decays to radially excited pseudoscalar $2S$ and $3S$ states. And the corresponding results of vector final states are shown in tables~\ref{tab:112p} and \ref{tab:113p}. From these tables, we can see that the partial decay widths for the $2S$ final states are about two orders of magnitude smaller than those of the $1S$ final states. The node structure of the $2S$ wave function is responsible for the small rate. When calculate the overlap integral of the wave functions, as there is no node for the initial wave functions, the positive part and the negative part of the final wave functions will give contributions which cancel each other out, resulting in a small decay width. As to the $3S$ states, there are even severe cancellation, which leads to the smaller branching ratios.

The corresponding results of non-leptonic decays with the final heavy meson being $D(2S)$ or $D(3S)$ are shown in tables \ref{tab:BD2sX} and \ref{tab:BD3sX}, respectively. And the results for the channels with $D^\ast(2S)$ or $D^\ast(3S)$ being the final meson are presented in tabels~\ref{tab:BDstar2sX} and \ref{tab:BDstar3sX}, respectively. Most of the branching ratios are of the order of $10^{-5}\sim 10^{-6}$, which are beyond the detection capability of experiments nowadays. However, there are also several channels, such as $B_c\to \eta(2S)\rho$ and $B_c\to B_s(2S)\pi$, whose branching ratios are up to $10^{-4}$. Besides, in table \ref{tab: compare2S of semileptonic} and \ref{tab: compare2S of nonleptonic}, we also compare our results of decay channels with $2S$ final states with those of other methods. One can see that for the $B_c$ decay channels, our results are close to those of other models, while for the $B_s$ decay channels, ours are several times smaller than those in Ref.~\cite{Faustov:2012mt}.

\begin{table}[htbp]
\centering
 \caption{\label{tab:002s} Branching ratios ($10^{-4}$) and partial widths ($10^{-16}$ GeV) of semi-leptonic decays $0^- \rightarrow 0^-(2S)l\nu$.}
 \begin{tabular}{cccccccc}
   \Xhline{1.5pt}
  Channels                          &width      &   Br       \\
  \midrule
$B^-\rightarrow D^{0}(2S)e\nu_e$      &$0.600^{+0.142}_{-0.115}$      & $1.50^{+0.35}_{-0.29}$  
\\
$B^-\rightarrow D^{0}(2S)\mu \nu_\mu$   &$0.597^{+0.140}_{-0.115}$    & $1.49^{+0.35}_{-0.29}$
\\
$B^-\rightarrow D^{0}(2S)\tau \nu_\tau$ &$0.0599^{+0.0131}_{-0.0127}$ &
$0.149^{+0.032}_{-0.032}$
\\
\hline
$\bar{B}^0\rightarrow D^{+}(2S)e\nu_e$      &$0.603^{+0.145}_{-0.119}$      &  $1.39^{+0.33}_{-0.27}$  
\\
$\bar{B}^0\rightarrow D^{+}(2S)\mu \nu_\mu$    &$0.599^{+0.143}_{-0.118}$   & $1.38^{+0.33}_{-0.27}$
\\
$\bar{B}^0\rightarrow D^{+}(2S)\tau \nu_\tau$  &$0.0586^{+0.0132}_{-0.0128}$   & $0.135^{+0.030}_{-0.029}$
\\
\hline
$\bar{B}_s^0\rightarrow D_s^{+}(2S)e\nu_e$     &$1.36^{+0.39}_{-0.32}$   & $3.14^{+0.89}_{-0.73}$ 
\\
$\bar{B}_s^0\rightarrow D_s^{+}(2S)\mu \nu_\mu$ &$1.36^{+0.39}_{-0.31}$  & $3.12^{+0.89}_{-0.72}$
\\
$\bar{B}_s^0\rightarrow D_s^{+}(2S)\tau \nu_\tau$ &$0.106^{+0.030}_{-0.024}$   & $0.244^{+0.068}_{-0.056}$ 
\\
\hline
$B_c^-\rightarrow \eta_c(2S)e\nu_e$      &$6.09^{+0.81}_{-0.78}$   & $4.96^{+0.62}_{-0.60}$ 
\\
$B_c^-\rightarrow \eta_c(2S)\mu \nu_\mu$  &$6.04^{+0.80}_{-0.78}$  &  $4.65^{+0.62}_{-0.60}$
\\
$B_c^-\rightarrow \eta_c(2S)\tau \nu_\tau$ &$0.333^{+0.053}_{-0.048}$   & $0.257^{+0.041}_{-0.037}$
\\
\hline
$B_c^+\rightarrow B^0(2S)e\nu_e$      &$0.0585^{+0.0064}_{-0.0058}$   & $0.0451^{+0.0049}_{-0.0045}$ 
\\
$B_c^+\rightarrow  B^0(2S)\mu \nu_\mu$  &$0.0433^{+0.0051}_{-0.0047}$  &  $0.0334^{+0.0039}_{-0.0026}$
\\
\hline
$B_c^+\rightarrow B_s(2S)e\nu_e$      &$0.0288^{+0.0049}_{-0.0042}$   & $0.222^{+0.037}_{-0.032}$  
\\
$B_c^+\rightarrow B_s(2S)\mu \nu_\mu$  &$0.0172^{+0.0031}_{-0.0027}$  &  $0.132^{+0.024}_{-0.021}$
\\
\bottomrule
 \end{tabular}
\end{table}

\begin{table}[htbp]
\centering
 \caption{\label{tab:003s} Branching ratios ($10^{-5}$) and partial widths ($10^{-18}$ GeV) of semi-leptonic decays $0^- \rightarrow 0^-(3S)l\nu$.}
 \begin{tabular}{ccccccc}
   \Xhline{1.5pt}
  Channels                          &width      &   Br                 \\
  \midrule
$B^-\rightarrow D^{0}(3S)e\nu_e$      &$3.61^{+1.00}_{-0.92}$      & $0.899^{+0.249}_{-0.229}$
\\
$B^-\rightarrow D^{0}(3S)\mu \nu_\mu$   &$3.58^{+1.01}_{-0.89}$    & $0.892^{+0.253}_{-0.224}$
\\
$B^-\rightarrow D^{0}(3S)\tau \nu_\tau$  &$0.0341^{+0.0115}_{-0.0096}$   & $0.00850^{+0.00286}_{-0.00241}$
 \\
\hline
$\bar{B}^0\rightarrow D^{-}(3S)e\nu_e$      &$3.78^{+1.04}_{-0.93}$      &  $0.873^{+0.240}_{-0.217}$
\\
$\bar{B}^0\rightarrow D^{-}(3S)\mu \nu_\mu$    &$3.72^{+1.05}_{-0.91}$   & $0.859^{+0.243}_{-0.209}$
\\
$\bar{B}^0\rightarrow D^{-}(3S)\tau \nu_\tau$  &$0.0341^{+0.0112}_{-0.0094}$   & $0.00788^{+0.00259}_{-0.00219}$
\\
\hline
$\bar{B}_s^0\rightarrow D_s^{-}(3S)e\nu_e$     &$11.0^{+2.8}_{-2.3}$   & $2.52^{+0.63}_{-0.55}$
\\
$\bar{B}_s^0\rightarrow D_s^{-}(3S)\mu \nu_\mu$ &$10.8^{+2.8}_{-2.3}$  & $2.48^{+0.63}_{-0.53}$
\\
$\bar{B}_s^0\rightarrow D_s^{-}(3S)\tau \nu_\tau$ &$0.0731^{+0.0205}_{-0.0172}$ &  $0.0168^{+0.0047}_{-0.0039}$
\\
\hline
$B_c^-\rightarrow \eta_c(3S)e\nu_e$      &$53.7^{+11.2}_{-9.9}$   & $4.14^{+0.86}_{-0.76}$
\\
$B_c^-\rightarrow \eta_c(3S)\mu \nu_\mu$  &$53.1^{+11.4}_{-9.6}$  &  $4.09^{+0.88}_{-0.74}$
\\
$B_c^-\rightarrow \eta_c(3S)\tau \nu_\tau$ &$0.558^{+0.140}_{-0.119}$  & $0.0430^{+0.0108}_{-0.0092}$
\\
\bottomrule
 \end{tabular}
\end{table}

\begin{table}[htbp]
\centering
 \caption{\label{tab:112p} Branching ratios ($10^{-4}$) and partial widths ($10^{-16}$ GeV) of semi-leptonic decays $0^- \rightarrow 1^-(2S)l\nu$.}

 \begin{tabular}{ccccccccc}
   \Xhline{1.5pt}
  Channels                          &width      &   Br            \\
  \midrule
$B^-\rightarrow D^{*0}(2S)e\nu_e$      &$0.850^{+0.337}_{-0.279}$      & $2.123^{+0.842}_{-0.696}$
\\
$B^-\rightarrow D^{*0}(2S)\mu \nu_\mu$   &$0.846^{+0.336}_{-0.278}$    & $2.113^{+0.837}_{-0.692}$
\\
$B^-\rightarrow D^{*0}(2S)\tau \nu_\tau$  &$0.062^{+0.007}_{-0.006}$    & $0.155^{+0.182}_{-0.160}$
\\
\hline
$\bar{B}^0\rightarrow D^{*-}(2S)e\nu_e$      &$0.840^{+0.339}_{-0.277}$      &  $1.942^{+0.782}_{-0.643}$
\\
$\bar{B}^0\rightarrow D^{*-}(2S)\mu \nu_\mu$    &$0.834^{+0.336}_{-0.277}$   & $1.932^{+0.777}_{-0.639}$
\\
$\bar{B}^0\rightarrow D^{*-}(2S)\tau \nu_\tau$ &$0.059^{+0.007}_{-0.003}$    & $0.137^{+0.164}_{-0.144}$
\\
\hline
$\bar{B}_s^0\rightarrow D_s^{*+}(2S)e\nu_e$     &$2.551^{+0.806}_{-0.667}$   & $5.873^{+1.852}_{-1.532}$ 
\\
$\bar{B}_s^0\rightarrow D_s^{*+}(2S)\mu \nu_\mu$ &$2.541^{+0.801}_{-0.663}$  & $5.842^{+1.843}_{-1.523}$
\\
$\bar{B}_s^0\rightarrow D_s^{*+}(2S)\tau \nu_\tau$ &$0.176^{+0.086}_{-0.073}$    & $0.405^{+0.200}_{-0.175}$ 
\\
\hline
$B_c^-\rightarrow \psi(2S)e\nu_e$      &$10.621^{+1.491}_{-1.401}$   & $8.182^{+1.151}_{-1.085}$ 
\\
$B_c^-\rightarrow \psi(2S)\mu \nu_\mu$  &$10.522^{+1.481}_{-1.391}$  &  $8.102^{+1.141}_{-1.071}$
\\
$B_c^-\rightarrow \psi(2S)\tau \nu_\tau$ &$0.531^{+0.085}_{-0.085}$    & $0.408^{+0.065}_{-0.065}$
\\
\hline
$B_c^+\rightarrow B^{*0}(2S)e\nu_e$      &$0.027^{+0.007}_{-0.006}$   & $0.021^{+0.006}_{-0.005}$
\\
$B_c^+\rightarrow B^{*0}(2S)\mu \nu_\mu$  &$0.018^{+0.005}_{-0.004}$  &  $0.014^{+0.004}_{-0.004}$
\\
\hline
$B_c^+\rightarrow B_s^*(2S)e\nu_e$      &$0.149^{+0.053}_{-0.043}$   & $0.142^{+0.041}_{-0.033}$
\\
$B_c^+\rightarrow B_s^*(2S)\mu \nu_\mu$  &$0.083^{+0.031}_{-0.025}$  &  $0.006^{+0.002}_{-0.002}$
\\
\bottomrule
 \end{tabular}
\end{table}

\begin{table}[htbp]
\centering
 \caption{\label{tab:113p} Branching ratios ($10^{-4}$) and partial widths ($10^{-17}$ GeV) of semi-leptonic decays $0^- \rightarrow 1^-(3S)l
 \nu$.}
 \begin{tabular}{ccccccc}
   \Xhline{1.5pt}
  Channels                          &width      &   Br                 \\
  \midrule
$B^-\rightarrow D^{*0}(3S)e\nu_e$      &$2.57^{+0.73}_{-0.25}$      & $0.641^{+0.183}_{-0.064}$
\\
$B^-\rightarrow D^{*0}(3S)\mu \nu_\mu$   &$2.54^{+0.72}_{-0.25}$    & $0.632^{+0.180}_{-0.063}$
\\
$B^-\rightarrow D^{*0}(3S)\tau \nu_\tau$  &$0.0302^{+0.0093}_{-0.0029}$   & $0.00752^{+0.00232}_{-0.00075}$
 \\
\hline
$\bar{B}^0\rightarrow D^{*+}(3S)e\nu_e$      &$2.57^{+0.44}_{-0.25}$      &  $0.594^{+0.103}_{-0.059}$
\\
$\bar{B}^0\rightarrow D^{*+}(3S)\mu \nu_\mu$    &$2.54^{+0.44}_{-0.25}$   & $0.586^{+0.103}_{-0.058}$
\\
$\bar{B}^0\rightarrow D^{*+}(3S)\tau \nu_\tau$  &$0.0292^{+0.0060}_{-0.0029}$   & $0.00673^{+0.00138}_{-0.00067}$
\\
\hline
$\bar{B}_s^0\rightarrow D_s^{*+}(3S)e\nu_e$     &$6.60^{+1.09}_{-0.92}$   & $1.52^{+0.25}_{-0.21}$
\\
$\bar{B}_s^0\rightarrow D_s^{*+}(3S)\mu \nu_\mu$ &$6.49^{+1.07}_{-0.90}$  & $1.49^{+0.25}_{-0.21}$
\\
$\bar{B}_s^0\rightarrow D_s^{*+}(3S)\tau \nu_\tau$ &$0.0357^{+0.0070}_{-0.0058}$ &  $0.00821^{+0.00161}_{-0.00132}$
\\
\hline
$B_c^-\rightarrow \psi(3S)e\nu_e$      &$14.1^{+2.4}_{-2.2}$   & $1.09^{+0.18}_{-0.16}$
\\
$B_c^-\rightarrow \psi(3S)\mu \nu_\mu$  &$13.9^{+2.4}_{-2.1}$  &  $1.07^{+0.18}_{-0.16}$
\\
$B_c^-\rightarrow \psi(3S)\tau \nu_\tau$ &$0.0131^{+0.0032}_{-0.0027}$  & $0.0101^{+0.0025}_{-0.0021}$
\\
\bottomrule
 \end{tabular}
\end{table}

\begin{table}[htbp]
\centering
 \caption{\label{tab:BD2sX} Branching ratios ($10^{-5}$) and partial widths ($10^{-17}$ GeV) of non-leptonic decays $0^- \rightarrow 0^-(2S)X$. }

 \begin{tabular}{cccccccccc}
   \Xhline{1.5pt}
  $\bar{B}^0\rightarrow D^+(2S)X^-$         & width   &   Br         \\
  \midrule
$\pi^-$                           &$1.98^{+0.51}_{-0.40}$       &  $4.58^{+1.19}_{-0.93}$              \\
$K^-$                              &$0.147^{+0.038}_{-0.030}$       &  $0.340^{+0.088}_{-0.069}$                \\
$\rho^-$                           &$4.46^{+1.15}_{-0.89}$      &  $10.3^{+2.7}_{-2.1}$              \\
$K^{*-}$                           &$0.241^{+0.062}_{-0.048}$      & $0.557^{+0.143}_{-0.112}$              \\
  \midrule
$B^-\rightarrow D^0(2S)X^-$            & width    &   Br            \\
  \hline
$\pi^-$                            &$1.96^{+0.49}_{-0.39}$       &$4.88^{+1.23}_{-0.96}$              \\
$K^-$                               &$0.145^{+0.037}_{-0.029}$       &$0.362^{+0.091}_{-0.071}$                \\
$\rho^-$                            &$4.41^{+1.11}_{-0.86}$      &$10.9^{+2.7}_{-2.1}$              \\
$K^{*-}$                           &$0.239^{+0.059}_{-0.048}$      &$0.595^{+0.150}_{-0.117}$              \\
  \midrule
$\bar{B}^0_s\rightarrow D_s^+(2S)X^-$      & width &  Br         \\
  \hline
$\pi^-$                             &$4.94^{+1.47}_{-1.16}$       &  $11.3^{+3.4}_{-2.7}$          \\
$K^-$                               &$0.364^{+0.108}_{-0.085}$       & $0.836^{+0.247}_{-0.196}$             \\
$\rho^-$                            &$10.9^{+3.2}_{-2.6}$      &  $24.9^{+7.4}_{-5.9}$        \\
$K^{*-}$                           &$0.585^{+0.173}_{-0.137}$      & $1.34^{+0.40}_{-0.32}$        \\
  \midrule
$B_c^- \rightarrow \eta_c(2S)X^-$      & width &  Br                   \\
  \hline
$\pi^-$                            &$21.0^{+2.7}_{-2.6}$       &  $16.7^{+2.1}_{-2.0}$   \\
$K^-$                               & $1.53^{+0.20}_{-0.19}$       &  $1.19^{+0.16}_{-0.15}$          \\
$\rho^-$                            &$46.3^{+6.2}_{-6.1}$      &  $35.6^{+4.8}_{-4.7}$          \\
$K^{*-}$                            &$2.48^{+0.33}_{-0.33}$      &  $1.91^{+0.26}_{-0.25}$   
\\
  \midrule
$B_c^+\rightarrow B_s^0(2S)X^+$          & width   &   Br           \\
  \hline
$\pi^+$                             &   $48.9^{+9.6}_{-8.4}$    & $38.4^{+7.4}_{-6.6}$              \\
  \midrule
$B_c+\rightarrow B^0(2S)X^+$         & width   &   Br           \\
  \hline
$\pi^+$                           &$6.78^{+0.88}_{-0.79}$       &  $5.22^{+0.68}_{-0.61}$              \\
  \bottomrule
 \end{tabular}
\end{table}

\begin{table}[htbp]
\centering
 \caption{\label{tab:BD3sX} Branching ratios ($10^{-6}$) and partial widths ($10^{-18}$ GeV) of non-leptonic decays  $0^- \rightarrow 0^-(3S)X$. }
 \begin{tabular}{ccccccc}
   \Xhline{1.5pt}
  $\bar{B}^0\rightarrow D^+(3S)X^-$        & width   &   Br             \\
  \midrule
$\pi^-$                           &$2.29^{+0.68}_{-0.54}$       & $5.31^{+1.56}_{-1.24}$              \\
$K^-$                             &$0.163^{+0.048}_{-0.039}$       &  $0.378^{+0.111}_{-0.089}$                \\
$\rho^-$                           &$4.49^{+1.32}_{-1.07}$      &  $10.4^{+3.1}_{-2.5}$              \\
$K^{*-}$                            &$0.231^{+0.068}_{-0.055}$      &  $0.533^{+0.157}_{-0.128}$              \\
  \midrule
$B^-\rightarrow D^0(3S)X^-$          & width    &   Br            \\
  \hline
$\pi^-$                             &$2.19^{+0.66}_{-0.52}$       & $5.48^{+1.63}_{-1.30}$              \\
$K^-$                            &$0.156^{+0.047}_{-0.038}$       &$0.392^{+0.116}_{-0.094}$                \\
$\rho^-$                            &$4.30^{+1.28}_{-1.05}$      &$10.78^{+3.2}_{-2.6}$              \\
$K^{*-}$                           &$0.221^{+0.066}_{-0.054}$      &$0.551^{+0.164}_{-0.135}$              \\
  \midrule
$\bar{B}_s^0\rightarrow D_s^+(3S)X^-$      & width   &   Br              \\
  \hline
$\pi^-$                            &$6.89^{+1.79}_{-1.47}$       &  $15.8^{+4.1}_{-3.4}$              \\
$K^-$                              &$0.488^{+0.127}_{-0.104}$       &  $1.12^{+0.29}_{-0.24}$                \\
$\rho^-$                            &$13.3^{+3.5}_{-2.9}$      & $30.6^{+8.0}_{-6.6}$              \\
$K^{*-}$                            &$0.683^{+0.180}_{-0.147}$      &  $1.56^{+0.41}_{-0.34}$              \\
  \midrule
$B_c^-\rightarrow \eta_c(3S)X^-$       & width   &   Br             \\
  \hline
$\pi^-$                            &$28.0^{+6.5}_{-5.5}$       &  $21.6^{+5.0}_{-4.2}$              \\
$K^-$                              &$1.99^{+0.47}_{-0.39}$       & $1.53^{+0.36}_{-0.30}$                \\
$\rho^-$                           &$55.8^{+13.4}_{-11.1}$      &  $42.9^{+10.2}_{-8.5}$              \\
$K^{*-}$                           &$2.89^{+0.70}_{-0.58}$      &  $2.25^{+0.54}_{-0.44}$
\\
  \bottomrule
 \end{tabular}
\end{table}

\begin{table}[htbp]
\centering
 \caption{\label{tab:BDstar2sX} Branching ratios ($10^{-5}$) and partial widths ($10^{-17}$ GeV) of non-leptonic decays  $0^- \rightarrow 1^-(2S)X$. }
 \begin{tabular}{cccccccccc}
   \Xhline{1.5pt}
  $\bar{B}^0\rightarrow D^{*+}(2S)X^-$        & width   &   Br                \\
  \midrule
$\pi^-$                            &$1.64^{+0.53}_{-0.40}$       & $3.80^{+1.22}_{-0.92}$              \\
$K^-$                             &$0.118^{+0.038}_{-0.029}$       &  $0.274^{+0.088}_{-0.067}$               \\
$\rho^-$                            &$1.15^{+0.35}_{-0.26}$      &  $2.67^{+0.81}_{-0.61}$              \\
$K^{*-}$                            & $0.0702^{+0.0216}_{-0.0163}$      &  $0.162^{+0.050}_{-0.038}$              \\
  \midrule
$B^-\rightarrow D^{*0}(2S)X^-$           & width    &   Br             \\
  \hline
$\pi^-$                             & $1.64^{+0.52}_{-0.40}$       &  $4.10^{+1.30}_{-0.99}$              \\
$K^-$                               & $0.118^{+0.038}_{-0.029}$       &  $0.295^{+0.095}_{-0.072}$                \\
$\rho^-$                            &$1.16^{+0.35}_{-0.26}$      &  $2.87^{+0.88}_{-0.66}$              \\
$K^{*-}$                            & $0.0698^{+0.0217}_{-0.0163}$      &  $0.173^{+0.054}_{-0.041}$              \\
  \midrule
$\bar{B}_s\rightarrow D^{*+}_s(2S)X^-$      & width &   Br          \\
  \hline
$\pi^-$                            & $4.71^{+1.48}_{-1.16}$       &  $10.8^{+3.39}_{-2.66}$           \\
$K^-$                              &$0.338^{+0.106}_{-0.083}$       &  $0.777^{+0.243}_{-0.190}$             \\
$\rho^-$                           &$2.06^{+0.72}_{-0.53}$      &  $4.75^{+1.64}_{-1.21}$          \\
$K^{*-}$                            &$0.145^{+0.053}_{-0.039}$      &  $0.332^{+0.122}_{-0.090}$             \\
  \midrule
$B_c^- \rightarrow \psi(2S)X^-$         & width   &   Br            \\
  \hline
$\pi^-$                            &$18.6^{+2.6}_{-2.3}$       &  $14.2^{+2.0}_{-1.8}$   \\
$K^-$                               &$1.32^{+0.19}_{-0.17}$       &  $1.02^{+0.14}_{-0.13}$     \\
$\rho^-$                           & $53.5^{+2.8}_{-2.2}$      &  $40.4^{+2.1}_{-1.7}$       \\
$K^{*-}$                            & $2.07^{+0.20}_{-0.16}$      &  $2.822^{+0.156}_{-0.126}$ \\
  \midrule
$B_c^+ \rightarrow B^{*0}_s(2S)X^+$        & width  &   Br             \\
  \hline
$\pi^+$                             &$12.1^{+3.6}_{-3.0}$       &  $9.23^{+2.82}_{-2.34}$              \\
  \midrule
$B_c^+\rightarrow B^{*0}(2S)X^+$        & width   &   Br             \\
  \hline
$\pi^+$                          &$1.69^{+3.7}_{-3.2}$       &  $1.31^{+0.28}_{-0.24}$              \\
  \bottomrule
 \end{tabular}
\end{table}

\begin{table}[htbp]
\centering
 \caption{\label{tab:BDstar3sX} Branching ratios ($10^{-5}$) and partial widths ($10^{-18}$ GeV) of non-leptonic decays  $0^- \rightarrow 1^-(3S)X$. }
 \begin{tabular}{ccccccc}
   \Xhline{1.5pt}
  $\bar{B}^0\rightarrow D^{*+}(3S)X^-$       & width   &   Br           \\
  \midrule
$\pi^-$                            &$6.43^{+1.09}_{-6.52}$       &  $1.48^{+0.25}_{-1.50}$              \\
$K^-$                              &$0.442^{+0.070}_{-0.451}$       &  $0.102^{+0.016}_{-0.104}$                \\
$\rho^-$                          & $4.46^{+1.91}_{-4.54}$      &  $1.03^{+0.44}_{-1.04}$              \\
$K^{*-}$                          &$0.306^{+0.134}_{-0.311}$      &  $0.0709^{+0.0310}_{-0.0719}$              \\
  \midrule
$B^-\rightarrow D^{*0}(3S)X^-$           & width    &   Br             \\
  \hline
$\pi^-$                            &$6.43^{+1.89}_{-6.47}$       &  $1.58^{+0.47}_{-1.61}$              \\
$K^-$                              &$0.442^{+0.119}_{-0.443}$       &  $0.108^{+0.030}_{-0.110}$                \\
$\rho^-$                            &$4.42^{+1.98}_{-4.49}$      &  $1.11^{+0.50}_{-1.12}$              \\
$K^{*-}$                            &$0.302^{+0.139}_{-0.308}$      &  $0.0753^{+0.0347}_{-0.0766}$              \\
  \midrule
$\bar{B}^0_s\rightarrow D^{*+}_s(3S)X^-$      & width   &   Br           \\
  \hline
$\pi^-$                             &$22.2^{+3.8}_{-3.2}$       &$5.10^{+0.87}_{-0.74}$              \\
$K^-$                               &$1.52^{+0.25}_{-0.22}$       &$0.349^{+0.059}_{-0.051}$                \\
$\rho^-$                           &$15.5^{+3.8}_{-3.0}$      &$3.55^{+0.88}_{-0.69}$              \\
$K^{*-}$                           &$1.09^{+0.27}_{-0.21}$      &  $0.251^{+0.061}_{-0.048}$              \\
  \midrule
$B_c^- \rightarrow \psi(3S)X^-$        & width   &   Br            \\
  \hline
$\pi^-$                            &$40.3^{+7.4}_{-6.4}$       &$3.11^{+0.57}_{-0.49}$              \\
$K^-$                              &$2.78^{+0.52}_{-0.44}$       &  $0.214^{+0.040}_{-0.034}$                \\
$\rho^-$                            &$43.6^{+10.1}_{-8.4}$      &  $3.35^{+0.78}_{-0.65}$              \\
$K^{*-}$                            &$2.98^{+0.72}_{-0.58}$      &  $0.229^{+0.055}_{-0.045}$
\\
  \bottomrule
 \end{tabular}
\end{table}

\begin{table}[htbp]
  \centering
   \caption{\label{tab: compare2S of semileptonic} Branching ratios ($10^{-4}$) of some  semi-leptonic decays compared compared between different models. }
   \begin{tabular}{cccccccc}
     \Xhline{1.5pt}
    Channels         &   Br   &Br(Wang) \cite{Wang:2012wk} &Br(Faustov) \cite{Faustov:2012mt}  \\
    \hline
  $\bar{B}_s^0\rightarrow D_s^{+}(2S)e\nu_e$       & $3.14^{+0.89}_{-0.73}$ & $9.9 \pm 2.7$  & $27 \pm 3$
  \\
  $\bar{B}_s^0\rightarrow D_s^{+}(2S)\tau \nu_\tau$   & $0.244^{+0.068}_{-0.056}$ & & $1.1 \pm 0.1$ 
  \\
  $B_c^+\rightarrow B^0(2S)e\nu_e$         & $0.0451^{+0.0049}_{-0.0045}$ & $0.0120\pm0.0060$ 
  \\
  $B_c^+\rightarrow B_s(2S)e\nu_e$         & $0.222^{+0.037}_{-0.032}$  & $0.037\pm0.016$ 
  \\
  $\bar{B}_s^0\rightarrow D_s^{*+}(2S)e\nu_e$        & $5.873^{+1.852}_{-1.532}$& & $38 \pm 4$ 
  \\
  $\bar{B}_s^0\rightarrow D_s^{*+}(2S)\tau \nu_\tau$    & $0.405^{+0.200}_{-0.175}$& & $1.5 \pm 0.2$ 
  \\
  \midrule
  Channels     &   Br  &Br(Chang) \cite{Chang:1992pt} &Br(Ebert) \cite{Ebert:2003cn} & Br(Liu)   \cite{Liu:1997hr}   \\
  \hline
  $B_c^-\rightarrow \eta_c(2S)e\nu_e$        & $4.96^{+0.62}_{-0.60}$ & 5.60   & 3.54  & 4.66
  \\
  $B_c^-\rightarrow \psi(2S)e\nu_e$      & $8.182^{+1.151}_{-1.085}$ & 11.2  & 3.38  & 1.43 
  \\
  \bottomrule
  \end{tabular}
\end{table}

\begin{table}[htbp]
  \centering
   \caption{\label{tab: compare2S of nonleptonic}  Branching ratios ($10^{-5}$) of some  non-leptonic decays compared compared between different models.} 

   \begin{tabular}{cccccccccc}
     \Xhline{1.5pt}
  $\bar{B}^0_s\rightarrow D_s^+(2S)X^-$       &  Br &Br(Faustov) \cite{Faustov:2012mt}         \\
    \hline
  $\pi^-$                        &  $11.3^{+3.4}_{-2.7}$    & 70           \\
  $K^-$                               & $0.836^{+0.247}_{-0.196}$   & 5            \\
  $\rho^-$                       &  $24.9^{+7.4}_{-5.9}$         &  170    \\
  $K^{*-}$                             & $1.34^{+0.40}_{-0.32}$     & 8       \\
    \midrule
  $B_c^-\rightarrow \eta_c(2S)X^-$      &   Br  &Br(Chang) \cite{Chang:1992pt} &Br(Ebert) \cite{Ebert:2003cn} & Br(Liu)   \cite{Liu:1997hr}            \\
    \hline
  $\pi^-$                                  &  $16.7^{+2.1}_{-2.0}$  &    21.4   & 15.7  & 20.5 \\
  $K^-$                                      &  $1.19^{+0.16}_{-0.15}$   &    1.60     & 1.15  & 1.52        \\
  $\rho^-$                                  &  $35.6^{+4.8}_{-4.7}$         &    49.8   & 33.0  & 48.5   \\
  $K^{*-}$                                  &  $1.91^{+0.26}_{-0.25}$   &    2.48  & 1.73  & 2.34  \\
 \midrule
  $\bar{B}_s^0\rightarrow D^{*+}_s(2S)X^-$      &   Br     & Br(Faustov) \cite{Faustov:2012mt}       \\
  \hline
$\pi^-$                             &  $10.8^{+3.39}_{-2.66}$    &  80        \\
$K^-$                                   &  $0.777^{+0.243}_{-0.190}$     & 6           \\
$\rho^-$                              &  $4.75^{+1.64}_{-1.21}$        &   22   \\
$K^{*-}$                                &  $0.332^{+0.122}_{-0.090}$     & 1.2        \\
  \midrule
$B_c^-\rightarrow \psi(2S)X^-$       &   Br  &Br(Chang) \cite{Chang:1992pt} &Br(Ebert) \cite{Ebert:2003cn} & Br(Liu)   \cite{Liu:1997hr} \\
\hline
$\pi^-$                                 &  $14.2^{+2.0}_{-1.8}$  &  20.1   & 9.92  & 5.85   \\
$K^-$                                     &  $1.02^{+0.14}_{-0.13}$  & 1.44   & 0.744  & 0.412     \\
$\rho^-$                              &  $40.4^{+2.1}_{-1.7}$    & 56.9    & 16.5  & 15.1     \\
$K^{*-}$                             &  $2.822^{+0.156}_{-0.126}$ & 3.04     & 0.909 & 0.751 \\
    \bottomrule
   \end{tabular}
  \end{table}

In conclusion, we have used the improved BS method to calculate some semi-leptonic and non-leptonic decays of $B$, $B_s$ and $B_c$ mesons. To check this method, we first studied the non-leptonic decay channels of $B$ and $B_s$ to the $1S$ final state. By comparing the results with data, we confirm the validity of this new method. Then we investigated the processes of $B$, $B_s$, and $B_c$ decaying to $2S$ or $3S$ final heavy mesons. Many of the branching ratios for these decay channels are two or three orders of magnitude smaller than those with the $1S$ final heavy meson, which can not reach the detection ability of current experiments. However, there are still many channels, such as the semi-leptonic decays $B \to D(2S)l\nu$, have the possibilities to be detected in current experiments.

\acknowledgments

This work was supported in part by the National Natural Science Foundation of China (NSFC) under Grant Nos.~11575048. We also thank the HEPC Studio at Physics School of Harbin Institute of Technology for access to computing resources through INSPUR-HPC@hepc.hit.edu.cn.

\paragraph{}

\bibliographystyle{JHEP}
\bibliography{bibfile2S3S}

\providecommand{\href}[2]{#2}\begingroup\raggedright\begin{thebibliography}{10}

\bibitem{Tanabashi:2018oca}
{\scshape Particle Data Group} collaboration, \emph{{Review of Particle
  Physics}}, \href{https://doi.org/10.1103/PhysRevD.98.030001}{\emph{Phys. Rev.
  D} {\bfseries 98} (2018) 030001}.

\bibitem{Lees:2012xj}
{\scshape BaBar} collaboration, \emph{{Evidence for an excess of $\bar{B} \to
  D^{(*)} \tau^-\bar{\nu}_\tau$ decays}},
  \href{https://doi.org/10.1103/PhysRevLett.109.101802}{\emph{Phys. Rev. Lett.}
  {\bfseries 109} (2012) 101802}
  [\href{https://arxiv.org/abs/1205.5442}{{\ttfamily 1205.5442}}].

\bibitem{Sato:2016svk}
{\scshape Belle} collaboration, \emph{{Measurement of the branching ratio of
  $\bar{B}^0 \rightarrow D^{*+} \tau^- \bar{\nu}_{\tau}$ relative to $\bar{B}^0
  \rightarrow D^{*+} \ell^- \bar{\nu}_{\ell}$ decays with a semileptonic
  tagging method}},
  \href{https://doi.org/10.1103/PhysRevD.94.072007}{\emph{Phys. Rev. D}
  {\bfseries 94} (2016) 072007}
  [\href{https://arxiv.org/abs/1607.07923}{{\ttfamily 1607.07923}}].

\bibitem{Aaij:2017uff}
{\scshape LHCb} collaboration, \emph{{Measurement of the ratio of the $B^0 \to
  D^{*-} \tau^+ \nu_{\tau}$ and $B^0 \to D^{*-} \mu^+ \nu_{\mu}$ branching
  fractions using three-prong $\tau$-lepton decays}},
  \href{https://doi.org/10.1103/PhysRevLett.120.171802}{\emph{Phys. Rev. Lett.}
  {\bfseries 120} (2018) 171802}
  [\href{https://arxiv.org/abs/1708.08856}{{\ttfamily 1708.08856}}].

\bibitem{581BARBAR:BDpi}
{\scshape BaBar} collaboration, \emph{{\color{red} Branching fraction
  measurement of $\bar{B}^{0} \rightarrow D^{(*)+} \pi^{-}$ and $B^{-}
  \rightarrow D^{(*) 0} \pi^{-}$ and isospin analysis of $\bar{B} \rightarrow
  D^{(*)} \pi$ decays}},
  \href{https://doi.org/10.1103/PhysRevD.75.031101}{\emph{Phys. Rev. D}
  {\bfseries 75} (2007) 031101}
  [\href{https://arxiv.org/abs/hep-ex/0610027}{{\ttfamily hep-ex/0610027}}].

\bibitem{699Belle:B+DK-}
{\scshape Belle} collaboration, \emph{{\color{red} Measurement of branching
  fraction ratios and CP asymmetries in $B^{\pm} \rightarrow D_{CP} K^{\pm}$}},
  \href{https://doi.org/10.1103/PhysRevD.68.051101}{\emph{Phys. Rev. D}
  {\bfseries 68} (2003) 051101}
  [\href{https://arxiv.org/abs/hep-ex/0304032}{{\ttfamily hep-ex/0304032}}].

\bibitem{744LHCb:BsDsKpi}
{\scshape LHCb} collaboration, \emph{{Measurements of the branching fractions
  of the decays $B^{0}_{s} \to D^{\mp}_{s} K^{\pm} $ and $B^{0}_{s} \to
  D^{-}_{s} \pi^{+}$}},
  \href{https://doi.org/10.1007/JHEP06(2012)115}{\emph{JHEP} {\bfseries 06}
  (2012) 115} [\href{https://arxiv.org/abs/1204.1237}{{\ttfamily 1204.1237}}].

\bibitem{Neubert:1997uc}
M.~Neubert and B.~Stech, \emph{{Nonleptonic weak decays of B mesons}},
  \href{https://arxiv.org/abs/hep-ph/9705292}{{\ttfamily hep-ph/9705292}}.

\bibitem{Beneke:2000ry}
M.~Beneke, G.~Buchalla, M.~Neubert and C.~T. Sachrajda, \emph{{QCD
  factorization for exclusive, nonleptonic B meson decays: General arguments
  and the case of heavy light final states}},
  \href{https://doi.org/10.1016/S0550-3213(00)00559-9}{\emph{Nucl. Phys. B}
  {\bfseries 591} (2000) 313}
  [\href{https://arxiv.org/abs/hep-ph/0006124}{{\ttfamily hep-ph/0006124}}].

\bibitem{Li:2008ts}
R.-H. Li, C.-D. Lu and H.~Zou, \emph{{The $B(B_{(s)}) \to D_{(s)} P$, $D_{(s)}
  V$, $D^*_{(s)} P$ and $D^*_{(s)} V$ decays in the perturbative QCD
  approach}}, \href{https://doi.org/10.1103/PhysRevD.78.014018}{\emph{Phys.
  Rev. D} {\bfseries 78} (2008) 014018}
  [\href{https://arxiv.org/abs/0803.1073}{{\ttfamily 0803.1073}}].

\bibitem{Li:2010bb}
G.~Li, F.-l. Shao and W.~Wang, \emph{{$B_s \to D_s(3040)$ form factors and
  $B_s$ decays into $D_s(3040)$}},
  \href{https://doi.org/10.1103/PhysRevD.82.094031}{\emph{Phys. Rev. D}
  {\bfseries 82} (2010) 094031}
  [\href{https://arxiv.org/abs/1008.3696}{{\ttfamily 1008.3696}}].

\bibitem{Li:2009wq}
R.-H. Li, C.-D. Lu and Y.-M. Wang, \emph{{\color{red} Exclusive $B_{s}$ decays
  to the charmed mesons $D_{s}^{+}(1968,2317)$ in the standard model}},
  \href{https://doi.org/10.1103/PhysRevD.80.014005}{\emph{Phys. Rev. D}
  {\bfseries 80} (2009) 014005}
  [\href{https://arxiv.org/abs/0905.3259}{{\ttfamily 0905.3259}}].

\bibitem{Chang:1992pt}
C.-H. Chang and Y.-Q. Chen, \emph{{The Decays of $B_{(c)}$ meson}},
  \href{https://doi.org/10.1103/PhysRevD.49.3399}{\emph{Phys. Rev. D}
  {\bfseries 49} (1994) 3399}.

\bibitem{Chang:2014jca}
C.~Chang, H.-F. Fu, G.-L. Wang and J.-M. Zhang, \emph{{Some of semileptonic and
  nonleptonic decays of $B_c$ meson in a Bethe-Salpeter relativistic quark
  model}}, \href{https://doi.org/10.1007/s11433-015-5671-x}{\emph{Sci. China
  Phys. Mech. Astron.} {\bfseries 58} (2015) 071001}
  [\href{https://arxiv.org/abs/1411.3428}{{\ttfamily 1411.3428}}].

\bibitem{Anisimov:1998xv}
A.~Anisimov, P.~Kulikov, I.~Narodetsky and K.~Ter-Martirosian, \emph{{Exclusive
  and inclusive decays of the $B_c$ meson in the light front ISGW model}},
  {\emph{Phys. Atom. Nucl.} {\bfseries 62} (1999) 1739}
  [\href{https://arxiv.org/abs/hep-ph/9809249}{{\ttfamily hep-ph/9809249}}].

\bibitem{Ebert:2003cn}
D.~Ebert, R.~Faustov and V.~Galkin, \emph{{Weak decays of the $B_c$ meson to
  charmonium and $D$ mesons in the relativistic quark model}},
  \href{https://doi.org/10.1103/PhysRevD.68.094020}{\emph{Phys. Rev. D}
  {\bfseries 68} (2003) 094020}
  [\href{https://arxiv.org/abs/hep-ph/0306306}{{\ttfamily hep-ph/0306306}}].

\bibitem{Ivanov:2005fd}
M.~A. Ivanov, J.~G. Korner and P.~Santorelli, \emph{{Semileptonic decays of
  $B_c$ mesons into charmonium states in a relativistic quark model}},
  \href{https://doi.org/10.1103/PhysRevD.75.019901}{\emph{Phys. Rev. D}
  {\bfseries 71} (2005) 094006}
  [\href{https://arxiv.org/abs/hep-ph/0501051}{{\ttfamily hep-ph/0501051}}].

\bibitem{Ivanov:2000aj}
M.~A. Ivanov, J.~Korner and P.~Santorelli, \emph{{The Semileptonic decays of
  the $B_c$ meson}},
  \href{https://doi.org/10.1103/PhysRevD.63.074010}{\emph{Phys. Rev. D}
  {\bfseries 63} (2001) 074010}
  [\href{https://arxiv.org/abs/hep-ph/0007169}{{\ttfamily hep-ph/0007169}}].

\bibitem{Ivanov:2006ni}
M.~A. Ivanov, J.~G. Korner and P.~Santorelli, \emph{{Exclusive semileptonic and
  nonleptonic decays of the $B_c$ meson}},
  \href{https://doi.org/10.1103/PhysRevD.73.054024}{\emph{Phys. Rev. D}
  {\bfseries 73} (2006) 054024}
  [\href{https://arxiv.org/abs/hep-ph/0602050}{{\ttfamily hep-ph/0602050}}].

\bibitem{Liu:1997hr}
J.-F. Liu and K.-T. Chao, \emph{{$B_c$ meson weak decays and CP violation}},
  \href{https://doi.org/10.1103/PhysRevD.56.4133}{\emph{Phys. Rev. D}
  {\bfseries 56} (1997) 4133}.

\bibitem{Choi:2002na}
{\scshape Belle} collaboration, \emph{{Observation of the $\eta_{c}(2 S)$ in
  exclusive $B \rightarrow K K_{S} K^{-} \pi^{+}$ decays}},
  \href{https://doi.org/10.1103/PhysRevLett.89.102001}{\emph{Phys. Rev. Lett.}
  {\bfseries 89} (2002) 102001}
  [\href{https://arxiv.org/abs/hep-ex/0206002}{{\ttfamily hep-ex/0206002}}].

\bibitem{Brodzicka:2007aa}
{\scshape Belle} collaboration, \emph{{Observation of a new $D_{(sJ)}$ meson in
  $B^+ \to \bar{D}^0 D^0 K^+$ decays}},
  \href{https://doi.org/10.1103/PhysRevLett.100.092001}{\emph{Phys. Rev. Lett.}
  {\bfseries 100} (2008) 092001}
  [\href{https://arxiv.org/abs/0707.3491}{{\ttfamily 0707.3491}}].

\bibitem{Matsuki:2006rz}
T.~Matsuki, T.~Morii and K.~Sudoh, \emph{{Radial Excitations of Heavy Mesons}},
  \href{https://doi.org/10.1140/epja/i2006-10287-1}{\emph{Eur. Phys. J. A}
  {\bfseries 31} (2007) 701}
  [\href{https://arxiv.org/abs/hep-ph/0610186}{{\ttfamily hep-ph/0610186}}].

\bibitem{Godfrey:2013aaa}
S.~Godfrey and I.~T. Jardine, \emph{{Nature of the $D_{s1}^*$(2710) and
  $D_{sJ}^*$(2860) mesons}},
  \href{https://doi.org/10.1103/PhysRevD.89.074023}{\emph{Phys. Rev. D}
  {\bfseries 89} (2014) 074023}
  [\href{https://arxiv.org/abs/1312.6181}{{\ttfamily 1312.6181}}].

\bibitem{Ebert:2009ua}
D.~Ebert, R.~Faustov and V.~Galkin, \emph{{Heavy-light meson spectroscopy and
  Regge trajectories in the relativistic quark model}},
  \href{https://doi.org/10.1140/epjc/s10052-010-1233-6}{\emph{Eur. Phys. J. C}
  {\bfseries 66} (2010) 197} [\href{https://arxiv.org/abs/0910.5612}{{\ttfamily
  0910.5612}}].

\bibitem{Wang:2013mml}
G.-L. Wang, Y.~Jiang, T.~Wang and W.-L. Ju, \emph{{The Properties of
  $D^{*}_{s1}(2700)^{+}$}},  \href{https://arxiv.org/abs/1305.4756}{{\ttfamily
  1305.4756}}.

\bibitem{Aaij:2019sqk}
{\scshape LHCb} collaboration, \emph{{Determination of quantum numbers for
  several excited charmed mesons observed in $B^- \to D^{*+} \pi^- \pi^-$
  decays}}, \href{https://doi.org/10.1103/PhysRevD.101.032005}{\emph{Phys. Rev.
  D} {\bfseries 101} (2020) 032005}
  [\href{https://arxiv.org/abs/1911.05957}{{\ttfamily 1911.05957}}].

\bibitem{Godfrey:1985xj}
S.~Godfrey and N.~Isgur, \emph{{Mesons in a Relativized Quark Model with
  Chromodynamics}}, \href{https://doi.org/10.1103/PhysRevD.32.189}{\emph{Phys.
  Rev. D} {\bfseries 32} (1985) 189}.

\bibitem{Aaij:2013sza}
{\scshape LHCb} collaboration, \emph{{Study of $D_J$ meson decays to
  $D^+\pi^-$, $D^0 \pi^+$ and $D^{*+}\pi^-$ final states in pp collision}},
  \href{https://doi.org/10.1007/JHEP09(2013)145}{\emph{JHEP} {\bfseries 09}
  (2013) 145} [\href{https://arxiv.org/abs/1307.4556}{{\ttfamily 1307.4556}}].

\bibitem{Abreu:1998vk}
{\scshape DELPHI} collaboration, \emph{{First evidence for a charm radial
  excitation, $D^{*\prime}$}},
  \href{https://doi.org/10.1016/S0370-2693(98)00346-3}{\emph{Phys. Lett. B}
  {\bfseries 426} (1998) 231}.

\bibitem{Aaij:2016fma}
{\scshape LHCb} collaboration, \emph{{Amplitude analysis of $B^{-} \to D^{+}
  \pi^{-} \pi^{-}$ decays}},
  \href{https://doi.org/10.1103/PhysRevD.94.072001}{\emph{Phys. Rev. D}
  {\bfseries 94} (2016) 072001}
  [\href{https://arxiv.org/abs/1608.01289}{{\ttfamily 1608.01289}}].

\bibitem{Geng:2018qrl}
Z.-K. Geng, T.~Wang, Y.~Jiang, G.~Li, X.-Z. Tan and G.-L. Wang,
  \emph{{Relativistic effects in the semileptonic $B_c$ decays to charmonium
  with the Bethe-Salpeter method}},
  \href{https://doi.org/10.1103/PhysRevD.99.013006}{\emph{Phys. Rev. D}
  {\bfseries 99} (2019) 013006}
  [\href{https://arxiv.org/abs/1809.02968}{{\ttfamily 1809.02968}}].

\bibitem{Zhou:2019stx}
T.~Zhou, T.-h. Wang, Y.~Jiang, X.-Z. Tan, G.~Li and G.-L. Wang,
  \emph{{Relativistic calculations of $R(D^{(*)})$, $R(D^{(*)}_s)$, $R(\eta_c)$
  and $R(J/\psi)$}},
  \href{https://doi.org/10.1142/S0217751X20500761}{\emph{Int. J. Mod. Phys. A}
  {\bfseries 35} (2020) 2050076}
  [\href{https://arxiv.org/abs/1910.06595}{{\ttfamily 1910.06595}}].

\bibitem{Fu:2011tn}
H.-F. Fu, Y.~Jiang, C.~Kim and G.-L. Wang, \emph{{Probing Non-leptonic Two-body
  Decays of $B_c$ meson}},
  \href{https://doi.org/10.1007/JHEP06(2011)015}{\emph{JHEP} {\bfseries 06}
  (2011) 015} [\href{https://arxiv.org/abs/1102.5399}{{\ttfamily 1102.5399}}].

\bibitem{Buchalla:1995vs}
G.~Buchalla, A.~J. Buras and M.~E. Lautenbacher, \emph{{Weak decays beyond
  leading logarithms}},
  \href{https://doi.org/10.1103/RevModPhys.68.1125}{\emph{Rev. Mod. Phys.}
  {\bfseries 68} (1996) 1125}
  [\href{https://arxiv.org/abs/hep-ph/9512380}{{\ttfamily hep-ph/9512380}}].

\bibitem{Kim:2003ny}
C.~S. Kim and G.-L. Wang, \emph{{Average kinetic energy of heavy quark
  $(\mu^2(\pi))$ inside heavy meson of $0^-$ state by Bethe-Salpeter method}},
  \href{https://doi.org/10.1016/j.physletb.2004.01.058,
  10.1016/j.physletb.2006.01.053}{\emph{Phys. Lett. B} {\bfseries 584} (2004)
  285} [\href{https://arxiv.org/abs/hep-ph/0309162}{{\ttfamily
  hep-ph/0309162}}].

\bibitem{Wang:2012cp}
T.~Wang, G.-L. Wang, Y.~Jiang and W.-L. Ju, \emph{{Electromagnetic Decay of
  $X(3872)$ as the $1{^1D_2}(2^{-+})$ charmonium}},
  \href{https://doi.org/10.1088/0954-3899/40/3/035003}{\emph{J. Phys. G}
  {\bfseries 40} (2013) 035003}
  [\href{https://arxiv.org/abs/1205.5725}{{\ttfamily 1205.5725}}].

\bibitem{Wang:2005qx}
G.-L. Wang, \emph{{Decay constants of heavy vector mesons in relativistic
  Bethe-Salpeter method}},
  \href{https://doi.org/10.1016/j.physletb.2005.12.005}{\emph{Phys. Lett. B}
  {\bfseries 633} (2006) 492}
  [\href{https://arxiv.org/abs/math-ph/0512009}{{\ttfamily math-ph/0512009}}].

\bibitem{Godfrey:2015dva}
S.~Godfrey and K.~Moats, \emph{{Properties of Excited Charm and Charm-Strange
  Mesons}}, \href{https://doi.org/10.1103/PhysRevD.93.034035}{\emph{Phys. Rev.
  D} {\bfseries 93} (2016) 034035}
  [\href{https://arxiv.org/abs/1510.08305}{{\ttfamily 1510.08305}}].

\bibitem{Barnes:2005pb}
T.~Barnes, S.~Godfrey and E.~Swanson, \emph{{Higher charmonia}},
  \href{https://doi.org/10.1103/PhysRevD.72.054026}{\emph{Phys. Rev. D}
  {\bfseries 72} (2005) 054026}
  [\href{https://arxiv.org/abs/hep-ph/0505002}{{\ttfamily hep-ph/0505002}}].

\bibitem{Wang:2014lml}
H.~Wang, Z.~Yan and J.~Ping, \emph{{Radially Excited States of $\eta_c$}},
  \href{https://doi.org/10.1140/epjc/s10052-015-3418-5}{\emph{Eur. Phys. J. C}
  {\bfseries 75} (2015) 196} [\href{https://arxiv.org/abs/1412.7068}{{\ttfamily
  1412.7068}}].

\bibitem{Aaij:2016utb}
{\scshape LHCb} collaboration, \emph{{Study of $D_{s J}^{(*)+}$ mesons decaying
  to $D^{*+} K_{S}^{0}$ and $D^{* 0} K^{+}$ final states}},
  \href{https://doi.org/10.1007/JHEP02(2016)133}{\emph{JHEP} {\bfseries 02}
  (2016) 133} [\href{https://arxiv.org/abs/1601.01495}{{\ttfamily
  1601.01495}}].

\bibitem{Aubert:2006jc}
{\scshape BaBar} collaboration, \emph{{Measurement of the absolute branching
  fractions $B$ to $D\pi$, $D^*\pi$, $D^{\pi}$ with a missing mass method}},
  \href{https://doi.org/10.1103/PhysRevD.74.111102}{\emph{Phys. Rev. D}
  {\bfseries 74} (2006) 111102}
  [\href{https://arxiv.org/abs/hep-ex/0609033}{{\ttfamily hep-ex/0609033}}].

\bibitem{Aaij:2017ryw}
{\scshape LHCb} collaboration, \emph{{Measurement of $CP$ observables in $B^\pm
  \to D^{(*)} K^\pm$ and $B^\pm \to D^{(*)} \pi^\pm$ decays}},
  \href{https://doi.org/10.1016/j.physletb.2017.11.070}{\emph{Phys. Lett. B}
  {\bfseries 777} (2018) 16}
  [\href{https://arxiv.org/abs/1708.06370}{{\ttfamily 1708.06370}}].

\bibitem{Abulencia:2006aa}
{\scshape CDF} collaboration, \emph{{Measurement of the Ratios of Branching
  Fractions $\mathcal{B}\left(B_{s}^{0} \rightarrow D_{s}^{-} \pi^{+} \pi^{+}
  \pi^{-}\right) / \mathcal{B}\left(B^{0} \rightarrow D^{-} \pi^{+} \pi^{+}
  \pi^{-}\right)$ and $\mathcal{B}\left(B_{s}^{0} \rightarrow D_{s}^{-}
  \pi^{+}\right) / \mathcal{B}\left(B^{0} \rightarrow D^{-} \pi^{+}\right)$}},
  \href{https://doi.org/10.1103/PhysRevLett.98.061802}{\emph{Phys. Rev. Lett.}
  {\bfseries 98} (2007) 061802}
  [\href{https://arxiv.org/abs/hep-ex/0610045}{{\ttfamily hep-ex/0610045}}].

\bibitem{Aaij:2015xga}
{\scshape LHCb} collaboration, \emph{{Measurement of the branching fraction
  ratio $\mathcal{B}(B_c^+ \rightarrow \psi(2S)\pi^+)/\mathcal{B}(B_c^+
  \rightarrow J/\psi \pi^+)$}},
  \href{https://doi.org/10.1103/PhysRevD.92.072007}{\emph{Phys. Rev. D}
  {\bfseries 92} (2015) 072007}
  [\href{https://arxiv.org/abs/1507.03516}{{\ttfamily 1507.03516}}].

\bibitem{Faustov:2012mt}
R.~Faustov and V.~Galkin, \emph{{Weak decays of $B_s$ mesons to $D_s$ mesons in
  the relativistic quark model}},
  \href{https://doi.org/10.1103/PhysRevD.87.034033}{\emph{Phys. Rev. D}
  {\bfseries 87} (2013) 034033}
  [\href{https://arxiv.org/abs/1212.3167}{{\ttfamily 1212.3167}}].

\bibitem{Wang:2012wk}
Z.-H. Wang, G.-L. Wang, J.-M. Zhang and T.-H. Wang, \emph{{The Productions and
  Strong Decays of $D_q(2S)$ and $B_q(2S)$}},
  \href{https://doi.org/10.1088/0954-3899/39/8/085006}{\emph{J. Phys. G}
  {\bfseries 39} (2012) 085006}
  [\href{https://arxiv.org/abs/1207.2528}{{\ttfamily 1207.2528}}].

\end{thebibliography}\endgroup

\end{document}